\documentstyle[12pt,epsfig,rotating] {article}
\begin{document}

\newcommand{\Z}{\mbox{$\mathrm{Z}$}}
\newcommand{\W}{\mbox{$\mathrm{W}$}}
\newcommand{\bZo}{{\bf \mbox{$\mathrm{Z}$}}}
\newcommand{\Zg}{\mbox{$\mathrm{Z}^{0}\gamma$}}
\newcommand{\ZZ}{\mbox{$\mathrm{Z}^{0}\mathrm{Z}^{0}$}}
\newcommand{\WW}{\mbox{$\mathrm{W}\mathrm{W}$}}
\newcommand{\Zs}{\mbox{$\mathrm{Z}^{*}$}}
\newcommand{\h}{\mbox{$\mathrm{h}^{0}$}}
\newcommand{\Ho}{\mbox{$\mathrm{H}$}}
\newcommand{\Hp}{\mbox{$\mathrm{H}^{+}$}}
\newcommand{\Hm}{\mbox{$\mathrm{H}^{-}$}}
\newcommand{\Hsm}{\mbox{$\mathrm{H}^{0}_{SM}$}}
\newcommand{\A}{\mbox{$\mathrm{A}^{0}$}}
\newcommand{\Hpm}{\mbox{$\mathrm{H}^{\pm}$}}
\newcommand{\X}{\mbox{${\tilde{\chi}^0}$}}
\newcommand{\ko}{\mbox{${\tilde{\chi}^0}$}}
\newcommand{\ee}{\mbox{$\mathrm{e}^{+}\mathrm{e}^{-}$}}
\newcommand{\bee}{\mbox{$\boldmath {\mathrm{e}^{+}\mathrm{e}^{-}} $}}
\newcommand{\mm}{\mbox{$\mu^{+}\mu^{-}$}}
\newcommand{\nn}{\mbox{$\nu \bar{\nu}$}}
\newcommand{\qq}{\mbox{$\mathrm{q} \bar{\mathrm{q}}$}}
\newcommand{\pb}{\mbox{$\mathrm{pb}^{-1}$}}
\newcommand{\ra}{\mbox{$\rightarrow$}}
\newcommand{\br}{\mbox{$\boldmath {\rightarrow}$}}
\newcommand{\tautau}{\mbox{$\tau^{+}\tau^{-}$}}
\newcommand{\ga}{\mbox{$\gamma$}}
\newcommand{\gamgam}{\mbox{$\gamma\gamma$}}
\newcommand{\tp}{\mbox{$\tau^+$}}
\newcommand{\tm}{\mbox{$\tau^-$}}
\newcommand{\tpm}{\mbox{$\tau^{\pm}$}}
\newcommand{\uu}{\mbox{$\mathrm{u} \bar{\mathrm{u}}$}}
\newcommand{\dd}{\mbox{$\mathrm{d} \bar{\mathrm{d}}$}}
\newcommand{\bb}{\mbox{$\mathrm{b} \bar{\mathrm{b}}$}}
\newcommand{\cc}{\mbox{$\mathrm{c} \bar{\mathrm{c}}$}}
\newcommand{\mumu}{\mbox{$\mu^+\mu^-$}}
\newcommand{\csbar}{\mbox{$\mathrm{c} \bar{\mathrm{s}}$}}
\newcommand{\cbars}{\mbox{$\bar{\mathrm{c}}\mathrm{s}$}}
\newcommand{\nunu}{\mbox{$\nu \bar{\nu}$}}
\newcommand{\nubar}{\mbox{$\bar{\nu}$}}
\newcommand{\mQ}{\mbox{$m_{\mathrm{Q}}$}}
\newcommand{\mZ}{\mbox{$m_{\mathrm{Z}}$}}
\newcommand{\mH}{\mbox{$m_{\mathrm{H}}$}}
\newcommand{\mHp}{\mbox{$m_{\mathrm{H}^+}$}}
\newcommand{\mh}{\mbox{$m_{\mathrm{h}}$}}
\newcommand{\mA}{\mbox{$m_{\mathrm{A}}$}}
\newcommand{\mHpm}{\mbox{$m_{\mathrm{H}^{\pm}}$}}
\newcommand{\mHsm}{\mbox{$m_{\mathrm{H}^0_{SM}}$}}
\newcommand{\mW}{\mbox{$m_{\mathrm{W}^{\pm}}$}}
\newcommand{\mt}{\mbox{$m_{\mathrm{t}}$}}
\newcommand{\mb}{\mbox{$m_{\mathrm{b}}$}}
\newcommand{\lpm}{\mbox{$\ell ^+ \ell^-$}}
\newcommand{\G}{\mbox{$\mathrm{GeV}$}}
\newcommand{\Gc}{\mbox{${\rm GeV}/c$}}
\newcommand{\Gcs}{\mbox{${\rm GeV}/c^2$}}
\newcommand{\Mcs}{\mbox{${\rm MeV}/c^2$}}
\newcommand{\sba}{\mbox{$\sin ^2 (\beta -\alpha)$}}
\newcommand{\cba}{\mbox{$\cos ^2 (\beta -\alpha)$}}
\newcommand{\tanb}{\mbox{$\tan \beta$}}
\newcommand{\sqrts}{\mbox{$\sqrt {s}$}}
\newcommand{\sqrtsp}{\mbox{$\sqrt {s'}$}}
\newcommand{\msusy}{\mbox{$M_{\rm SUSY}$}}
\newcommand{\mg}{\mbox{$m_{\tilde{\rm g}}$}}
\begin{titlepage}
\begin{center}
{\Large EUROPEAN ORGANIZATION FOR NUCLEAR RESEARCH}
\end{center}
%\bigskip
\begin{flushright}
  LHWG Note 2001-06 \\
  ALEPH 2001-36 CONF 2001-056 \\
  DELPHI 2001-116 CONF 540\\
  L3 note 2702\\
  OPAL Technical Note TN694\\
  {\today}
\end{flushright}
%\bigskip
\begin{center}{\Large \bf Searches for Invisible Higgs bosons:
Preliminary combined  
results using LEP data collected at energies up to 209~GeV}
\end{center}
\begin{center}
      {\Large  ALEPH, DELPHI, L3 and OPAL Collaborations}\\
      \bigskip
      {\large The LEP working group for Higgs boson searches
%\footnote{
%      Contributions from
%       ALEPH: P. McNamara, P. Teixeira-Dias, E. Locci, P. Colas, A. Tilquin, M. Kado;
%      DELPHI: S. Andringa, T. Fragat, P. Lutz, C. Martinez-Rivero, W. Murray, 
%              A. Read, V. Ruhlmann-Kleider, A. Sopczak, M. Stanitzki;
%      L3: M. Felcini, I. Fisk, P.Garcia-Abia, A. Holzner, A. Rosca, C. Tully, A. Raspereza;
%      OPAL: P. Bock, P. Igo-Kemenes, K. Hoffman, D. Horv\'ath,
%         T. Junk, I. Nakamura, A.N. Okpara, M. Oreglia, K. Nagai, A. Quadt, S. Yamashita;
%      Theory: M. Carena, S. Heinemeyer, P. Janot, C. Wagner, G. Weiglein.
%}
}
\end{center}
\bigskip
\begin{center}{\Large  Abstract}\end{center}
In the year 2000 the four LEP experiments have collected 
data at energies between 200 and 209 GeV, for approximately 868~\pb\ 
integrated luminosity.
The LEP working group for Higgs boson searches has combined these data 
with earlier data sets collected  at lower centre-of-mass energies
to search for a  neutral CP-even Higgs boson,
produced at the Standard Model rate,  decaying  into ``invisible"
particles.
No statistically
significant 
excess has been observed when compared to the Standard Model
background prediction, and assuming that the Higgs boson decays only into 
such states a  lower bound has been set on its mass at 
 95\% confidence level of 114.4~\Gcs.

\begin{center}
To be submitted to EPS'01 in Budapest and LP'01 in Rome
\end{center}

\begin{center}
ALL RESULTS QUOTED IN THIS NOTE ARE PRELIMINARY\\
\end{center}
\bigskip
\end{titlepage}
\section{Introduction}
We present combined results from the
ALEPH, DELPHI, L3 and OPAL Collaborations on searches for the 
 neutral Higgs bosons 
decaying into ``invisible" particles such as neutralinos~\cite{neutralinos} or majorons~\cite{majorons}. 
The results are obtained by   
combining the data collected in the year 2000 at 
centre-of-mass energies between 202 and 209~GeV with earlier data collected at
lower energies~\cite{adlo-cernep}. The new data represent an
integrated luminosity of approximately 868~\pb\ in total.

Unless explicitly specified, all cross-sections, branching ratios 
and many other physics quantities which are 
used in this combination of data, are calculated within HZHA~\cite{hzha}. 

Each experiment has generated
Monte Carlo event samples for the Higgs boson signal and the various 
background processes, typically,
at 202, 204, 206, 208 and 210~GeV energies. Cross-sections, branching ratios, 
distributions of the reconstructed mass and other 
discriminating variables relevant to the combination have been interpolated 
to energies which 
correspond to the data sets. In this procedure special care has been taken 
to the regions of kinematic cutoff where the signal and
background distributions vary rapidly. It has been established that the 
interpolation procedure does not add significantly
to the final systematic uncertainties.

The statistical procedure adopted for the combination of the data and the precise definition of the
confidence levels $CL_b,~CL_{s+b},~CL_s$ by which the search results 
are expressed, follow our  usual definitions\cite{adlo-cernep,stat}.

%%%%%%%%%%%%%%%%%%%%%%%%%%%%%%%%%%%%%%%%%%%%%%%%%%%%%%%%%%%%%%%%%%%%%%%%%%%%%%%%%%%%%%%%%%%%%%%%%%%%%%%%%%%
%
%%%%%%%%%%%%%%%%%%%%%%%%%%%%%%%%%%%%%%%%%%%%%%%%%%%%%%%%%%%%%%%%%%%%%%%%%%%%%%%
%
\section{Combined search for `invisible' Higgs boson decays}
At LEP the SM Higgs boson is expected
to be produced mainly via the Higgs-strahlung process \ee\ra~HZ, while
contributions from the WW\ra~H fusion channel, \ee\ra~H$\nu_e\bar{\nu_e}$, are typically 
below 10\%. 
The  Higgs boson may not make is presence obvious. 
For instance, in supersymmetric theories,
depending upon the parameters,
the decay of a Higgs boson into neutralinos might dominate.
The Higgs boson could then be invisible at LEP.
Majoron models can also produce
dominantly invisible  decay modes. However, if
the Higgs boson  is produced through the  Higgs-strahlung process,
the Z can be detected, and the presence of the Higgs
boson inferred.
This production processes is assumed here.

The four LEP collaborations  performed searches for
acoplanar jets  (H\ra invisible)(Z\ra\qq) or leptons 
 (H\ra invisible)(Z\ra ll).
%This is the first time such a combination has been made of the results
%of all four LEP experiments in this channel, and it  includes
%the  data collected in the year 2000 at centre-of-mass energies up to 209~GeV.

%All four LEP experiments, 
%were included in  this combination with  data from the 2000 LEP run.

The analysis procedures of the four LEP experiments producing the inputs for
the present  combination are described in individual
documents~\cite{hinvis-aleph, hinvis-delphi,hinvis-l3,hinvis-opal}; 
we merely summarise the results in Table~\ref{table-h-invis-input}. 
%This represents the full datasets of all the experiments.
For  DELPHI, the channel (Z\ra ll) includes the $\tau\tau$ decay mode
of the Z boson, which is not used by the other experiments.

%
%A plot of the candidate masses can be seen in Figure~\ref{mass_inv}.
%
%The figure has been obtained with the 
%supplementary requirement that the
%contributions from the four experiments (selecting the most signal-like set of events) be roughly
%equal. Since all events enter with equal weight, such a distribution does not reflect 
%for example differences in mass resolutions, signal sensitivities and background rates, 
%which characterise the various search channels and individual experiments.
%
%
%%
%\begin{figure}[htb]
%\begin{center}
%{\large INVISIBLE HIGGS - PRELIMINARY} \\
%\vspace{0.1cm}
%\epsfig{figure=lep_invis_200.eps,width=0.6\textwidth}
%\caption[]{\small \it 
%Missing mass for candidate events from the invisible Higgs
%boson final state search.
%The ALEPH analysis is mass dependent; the data for \mh=100~\Gcs
%has been used in this plot.
%\label{mass_inv}}
%\end{center}
%\end{figure}
%%
%%%
%
%\clearpage

\begin{table}[htb]
\begin{center}
\begin{tabular}{||l||c|c|c|c||}
\hline\hline
Experiment:                     & ALEPH & DELPHI & L3  & OPAL \\
\hline\hline
Integrated luminosity in 2000 (\pb):   & 215.6    &   225.1      &    217.3    & 210 \\
\phantom{.....}Backg. predicted / Evts. observed                                  &&&&  \\
\phantom{..........}Acoplanar jets:     & 8.17 / 8 &  35.9 / 30    &   56.2 / 50    & 61.8 / 47      \\ 
\phantom{..........}Acoplanar leptons:  & 6.7 /7  &   21.7 / 13    &   5.9 / 9    &    --          \\
\hline\hline
Events in all channels                  &  14.9 /15   &   57.6 / 43     &    62.1 / 59     & 61.8 / 47  \\
\hline
Median 95\% CL Limit (\Gcs)  :  &  112.6/111.8  &   110.7/110.7  &    110.2/110.1    & 107.4$^{*}$/108.5\\
Observed 95\% CL Limit (\Gcs)       :  &  114.1/113.1  &   113.0/113.0   &    107.6/107.5    & 107.0/107.4 \\
\hline\hline
\end{tabular}
\end{center}
\caption{\small\it Information related to the searches of the four LEP experiments for `invisible' Higgs boson decays
at energies between 200 and 209~GeV (year 2000 data), with a Higgs boson mass
at 110~\Gcs, if relevant.
All limits include the previous years' data.
The ALEPH confidence level estimator is different from that employed here.
Using the same technique we find an observed upper limit of 114.0 for the 
ALEPH data, which is close to the 114.1~\Gcs which ALEPH report.
(*) OPAL quotes the average instead of the median expected limit
\label{table-h-invis-input}}
\end{table}

The large spread in the numbers of selected candidates reflects substantial
differences in the selection methods and optimisation procedures. 
For instance, ALEPH and L3 use a sliding analysis technique. Only
the  relevant candidates for a  Higgs mass hypothesis are reported. 
For table~\ref{table-h-invis-input} the Higgs boson mass was taken as 110~\Gcs.
%All events which make an entry to 
%Table~\ref{table-h-invis-input} are used below in the calculation of
%confidence levels and in the limit setting procedure.
%

%
\begin{figure}[htb]
\begin{center}
\epsfig{figure=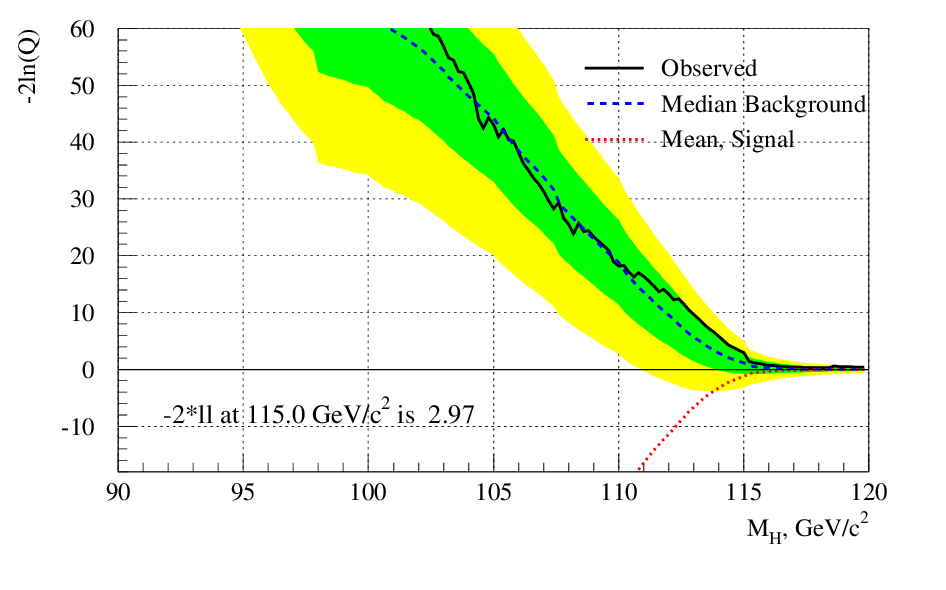,width=0.75\textwidth}
\vspace*{-0.8cm}
\caption[]{\small \it 
The distribution of -2*ln(Q) for the combined search for
$h^{0} \rightarrow invisible$\ by LEP.
\label{ll_inv}}
\end{center}
\end{figure}

The test-statistic as a function of the  mass \mH,
computed for the observed results, is shown in Figure~\ref{ll_inv}. 
In the presence  of a signal it should have a minimum near the
true Higgs boson mass. 
A negative value would
indicate a  preference for the signal hypothesis and the more negative 
the value the more significant the signal. The full-line curve representing
the observation 
is in agreement with the dashed line representing the background 
hypothesis, and deviates from the dotted curves
which represent the most likely signal + background situation. 
In fact, in the region 112 to 115~\Gcs
there is a deficit of events
correspondig to around  1.5 standard deviations from the background
expectation, which means that
the observed limit will be somewhat stronger than expected.

\begin{figure}[htb]
\begin{center}
\epsfig{figure=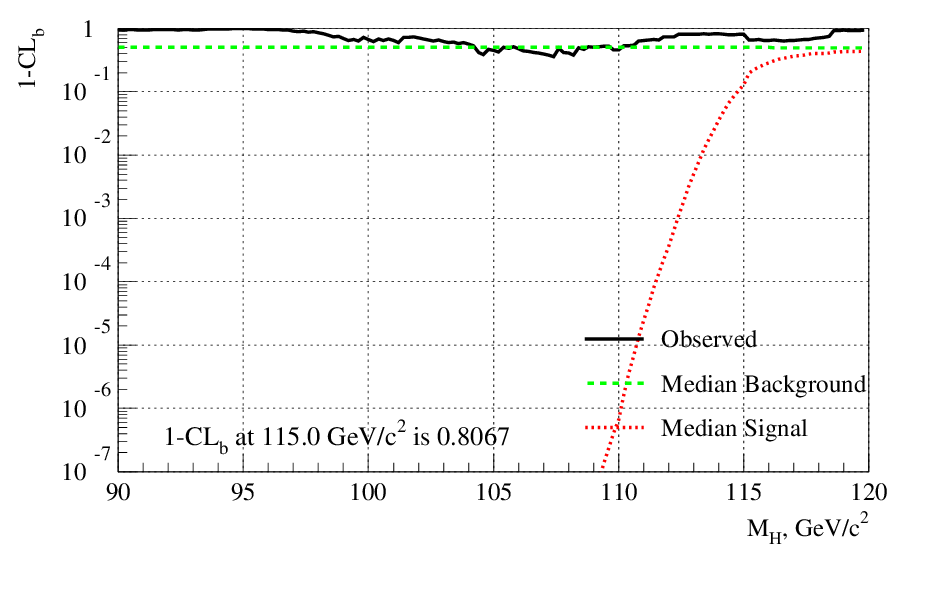,width=0.75\textwidth}
\vspace*{-0.8cm}
\caption[]{\small \it 
The value of $1-CL_{b}$ for the LEP combined data in the
$h^{0} \rightarrow invisible$
 search.
There is a no suggestion of any signal-like excess. 
%The step nature of the result comes because ALEPH and L3 used a
%sliding window analysis, and report relevant candidates in a discretised way.
%There is a 'knee' in the curve of signal expectations; below 98~\Gcs\ the
% ALEPH results from lower energies contribute.

\label{clb_inv}}

\end{center}
\end{figure}
The compatibility with background of the result is given by $1-CL_b$, 
which is plotted as a function of \mH\ 
in Figure~\ref{clb_inv}.
%Values of $1-CL_b$ below $5.7\times 10^{-7}$,
%indicated by the
%horizontal full line, corresponding to a 5 standard deviation
%fluctuation of the background, are considered to be in the discovery region.
The dotted line shows the  expectation in the presence of a signal;
%of true mass \mH;
 its crossing with the $5\sigma$ line at
109.5~\Gcs\ indicates the range of sensitivity of the  data to a discovery.
As expected, there is no suggestion of any signal.

\begin{figure}[htb]
\begin{center}
\epsfig{figure=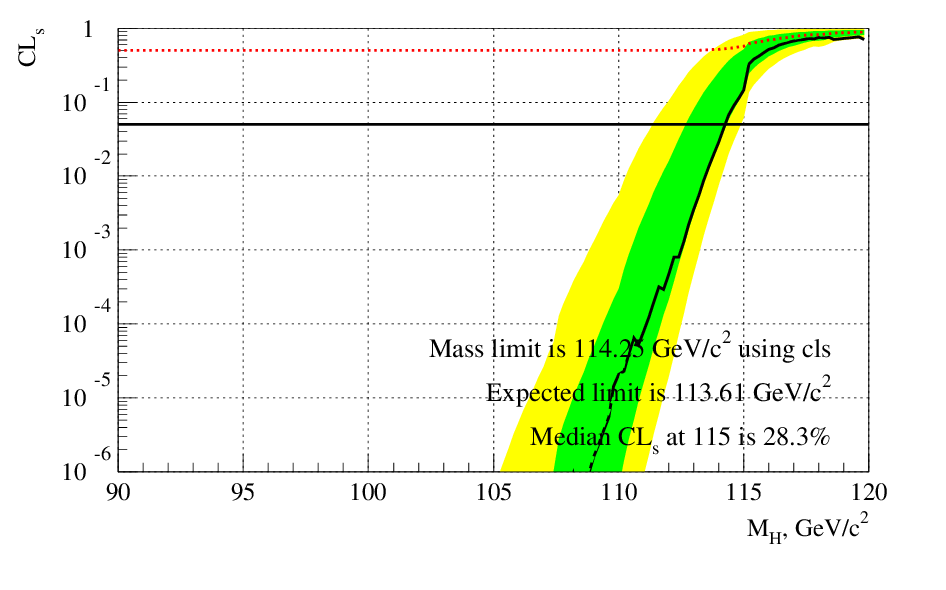,width=0.75\textwidth}
\vspace*{-0.8cm}
\caption[]{\small \it 
The value of $CL_{s}$ for the LEP combined data in the $h^{0} \rightarrow invisible$
 search. The observed limit of 114.4~\Gcs\ for a 100\% branching
ratio into invisible modes exceeds somewhat the expected limit of 113.6~\Gcs. 
\label{cls_inv}}
\end{center}
\end{figure}
A 95\% confidence level lower limit on the Higgs boson 
mass may be set by identifying the mass region where $CL_s < 0.05$, as  
shown in Figure~\ref{cls_inv}.
The $CL_s$ curve shown gives the limit on
$m_H$ assuming a 100\% branching ratio into invisible modes.
The median limit expected in the absence of a signal is 113.5~\Gcs\ and the
limit observed by combining the
LEP data is 114.4~\Gcs. The inclusion of systematic errors, which 
have been neglected, is expected to
reduce this lower bound by approximately 100~\Mcs.

%The excluded region in the $M_H$ v $Br(H\rightarrow invisible)$ plane is shown in
%Figure~\ref{2d_inv}.
%For all  masses shown  the exclusion is somewhat stronger than expected.
%The combination has not been performed for masses below 90~\Gcs.

%%
%\begin{figure}[htb]
%\begin{center}
%\epsfig{figure=inv2d_lep.eps,width=0.75\textwidth}
%\vspace*{-0.8cm}
%\caption[]{\small \it 
%The region excluded by the A,D,O result in the $h^{0} \rightarrow invisible$
% search. The filled area corresponds to a 95\% exclusion; 90 and 99\%
%exclusion contours are shown for comparison, and the results are similar.
%\label{2d_inv}}
%\end{center}
%\end{figure}
%
%

\begin{figure}[htb]
\begin{center}
\epsfig{figure=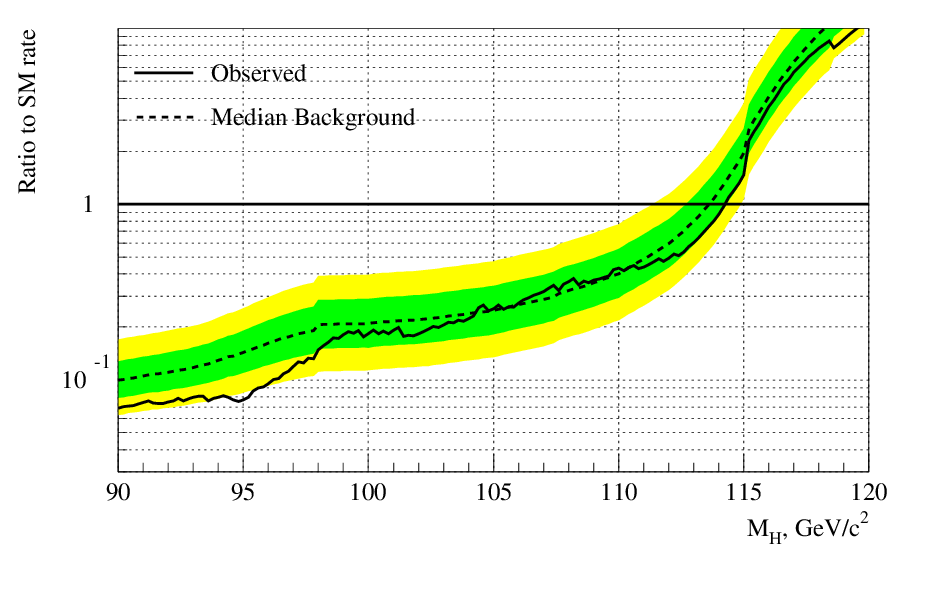,width=0.75\textwidth}
\vspace*{-0.8cm}
\caption[]{\small \it 
The region excluded by the combined LEP results in the $h^{0} \rightarrow invisible$
 search. The 95\% CL upper limit on, $\xi^2$, the production rate as a
fraction of the
Standard Model total rate,  is shown, together
with the expected range assuming there is no signal.
\label{fi:lim}}
\end{center}
\end{figure}

\begin{figure}[htb]
\begin{center}
\epsfig{figure=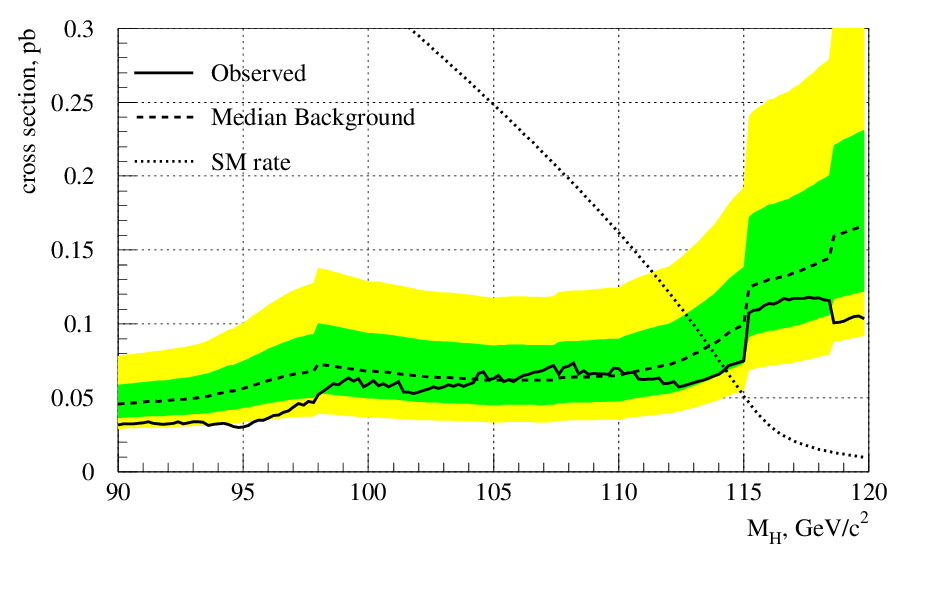,width=0.75\textwidth}
\vspace*{-0.8cm}
\caption[]{\small \it 
The upper limit on the invisible Higgs boson production cross-section.
The cross-section is quoted at  206.0~GeV; data at other energies
are scaled to that value using the Standard Model relative cross-sections.
\label{fi:lims}}
\end{center}
\end{figure}

The upper limit on the rate of $H \rightarrow invisible$ s a function
of $m_H$ is shown in Figure~\ref{fi:lim}. The scale is:

\begin{equation}
   \xi^2 = \frac{\sigma_{HZ}}{\sigma_{HZ}^{Standard Model}} Br(H \rightarrow invisible)
\end{equation}

 as a fraction of the
rate expected from a Standard Model H decaying 100\% invisibly.
The same information is shown in figure~\ref{fi:lims} but as a limit
on the cross-section at 206~GeV.
There is a somewhat stronger exclusion than expected around the $Z$ 
region, due to a deficit of candidates.


\clearpage

%%%%%%%%%%%%%%%%%%  END Hinv  %%%%%%%%%%%%%%%%%%%%%%%%%%%%%%%%%%

%
\section{Summary}
The LEP working group for Higgs boson searches has updated its previous
combined limit for the mass of a Higgs boson decaying  invisibly,
 including
the data collected in the year 2000 at energies between 200 and 209~GeV, 
for a total integrated luminosity of
approximately 868~\pb. In the absence of a statistically
significant excess in the data,
a new lower bound of 114.4~\Gcs\ has been obtained at the 95\% confidence level
and assuming the Standard Model production cross-section and that the
Higgs boson exclusively decays invisibly.

ALL THE RESULTS QUOTED IN THIS NOTE ARE PRELIMINARY.
%\newpage

%%%%%%%%%%%%%%%%%%%%%%%%%%%%%%%%%%%%%%%%%%%%%%%%%%%%%%%%%%%%%%%%%%%%%%%%%%%%%

\end{document}